\documentclass[prl,preprint,showpacs,showkeys]{revtex4}

\usepackage{graphics}
\usepackage{graphicx}
\usepackage{keyval}
\usepackage{float}
\usepackage{epstopdf}

\newcommand{\be}{\begin{equation}}
\newcommand{\ee}[1]{\label{#1} \end{equation}}
\newcommand{\ba}{\begin{eqnarray}}
\newcommand{\ea}[1]{\label{#1} \end{eqnarray}}

\newcommand{\td}[2]{ \frac{d#1}{d#2} }

\newcommand{\phott}{ \frac{dN}{k_{\perp}dk_{\perp}d\eta} }

\begin{document}


\title{{\bf Illusory Flow in Radiation from Accelerating Charge}}

\author{{ Tam\'as S. Bir\'o} }
	\affiliation{ MTA WIGNER Research Centre for Physics, RMI, Budapest, Hungary}

\author{{ Zsuzsanna Szendi} }
	\affiliation{ MTA WIGNER Research Centre for Physics, RMI, Budapest, Hungary}

\author{{Zsolt Schram} }
	\affiliation{Department of Theoretical Physics, University of Debrecen, Debrecen, Hungary}
        \affiliation{MTA-DE Particle Physics Research Group, Debrecen, Hungary}
\date{\today}

\vspace{12mm}

\begin{abstract}
In this paper we analyze the classical electromagnetic radiation of an accelerating point charge
moving on a straight line trajectory. 
Depending on the duration of accelerations, 
rapidity distributions of photons emerge, resembling the ones 
obtained in the framework of hydrodynamical models by Landau or Bjorken.
Detectable differences between our approach and spectra obtained from  hydrodynamical models
occur at high transverse momenta and are due to interference.
\end{abstract}



\maketitle


\section{Introduction}


Thermal and flow models accompany the history of heavy-ion collisions from the
beginnings. The idea of interpreting the spectra of newly produced hadrons
in high energy collisions in terms of a temperature dates back to Rolf Hagedorn
\cite{HAGEDORN,Hagedorn2}. It was even a pre-QCD observation that this temperature,
a measure for a presumably equipartitioned energy per particle, $T \sim E/N$,
cannot grow beyond limits: a limiting (maximal) temperature has been
understood in terms of an exponentially growing mass spectrum of heavy meson 
(and later also baryon) resonances \cite{BIRO+PESHIER,Broniowski}. 
Later this temperature, $T_H \approx 165$ MeV, has been identified as a color deconfinement
temperature beyond which hadrons gradually cease to exist and a special form of quark matter,
the quark-gluon plasma forms. This expectation has been a drive behind decades
of heavy ion experiments \cite{HEAVYION:REVIEW,Muller+Schafer,Muller+Bass,BMuller,Fries+Muller}. 

A hot fireball also expands, especially in vacuum. Theoretical hydrodynamical
solutions describing either locally isotropic or elongated fireballs were suggested by
Landau \cite{LANDAU:HYDRO} and Khalatnikov \cite{Khalatnikov} on one hand, and by Hwa \cite{HWA} and 
later Bjorken \cite{BJORKEN} on the other hand. Numerical hydrodynamical models
followed starting already in the Bevalac era at $\sqrt{s_{NN}}\sim 1-2$ GeV 
\cite{OLD:FRANKFURT:Buchwald,OLD:FRANKFURT:Rentzsch,OLD:FRANKFURT:Waldhauser,OLD:FRANKFURT:Dimitru,OLD:FRANKFURT:Gridnev},
and such efforts persisted until today's RHIC and LHC experiments at a much higher bombarding energy
\cite{Heinz2,Heinz,Heinz1,Romatschke,Rischke,Bouras,Schenke,Muronga,Molnar}.

Surprisingly, with the advent of LHC experiments at much higher energy
than applied in the 80-s, also some opinions emerge about producing
a quark-gluon plasma even in proton-proton collisions \cite{CMS}.
The overwhelming success of the hydrodynamical and thermal approach
for reconstructing particle spectra in the soft QCD regime makes us
wonder what the reason can be behind of it. Is it simply a maximal entropy state
in the information theory sense after averaging over so many elementary
events? It would explain thermal features, but not a collective flow.


At the LHC energies in a proton-proton collision the reaction
zone is very energetic, but the volume and the time for making a
(near-) equilibrium state is missing. QCD based and field theoretical
calculations should reveal the mechanism of very fast entropy
production in the early phase of high energy collisions.
It turned out earlier that a candidate mechanism might be realized
by the chaotic dynamics of classical Yang-Mills fields
\cite{Muller,Trayanov,ChaosBook,Strickland} or by other nonlinear
plasma instabilities \cite{Mrowczynski}. All such searches for a
"collectivizing" mechanism rely on 
the infrared sector of quantum field theory, the essential dynamics
being of classical nature.


The Unruh effect, known since the mid seventies, 
fits in this line  \cite{UNRUH,HAWKING}. 
Here a single frequency radiation seen by
a constantly accelerating observer occurs as a thermal radiation
exactly following Planck's law. This apparent temperature
does not stem from a  detailed and long standing energy balance
with a heat bath, but is a consequence of the continously changing
Doppler red-shift. Based upon this effect even a single point charge,
accelerated on a straight line, produces a radiation pattern of photons,
which contains an exponential factor in the yield, relating the
absolute temperature-like parameter to the acceleration of the source.
We have recently studied the possibility of such a pseudothermal effect
in relation to gamma spectra obtained in RHIC experiment
\cite{Biro+Schram+Gyulassy}.

In this paper we demonstrate that not only a temperature-like effect
shows up in this scenario, but a hydrodynamical flow can easily be
fitted to the classical radiation pattern as well. The J\"uttner distribution,
containing a collective flow field, $u_i(x)$ in the factor $\exp(-u_ip^i/T)$,
occurs under an integral for an everlasting constant acceleration.
In our calculations for finite time accelerations we obtain a different pattern.
The rapidity distributions of the photons at different
transverse momenta, $k_{\perp}$, resemble the plateau behavior in 
the differential rapidity, $dN/d\eta$, for long
enough acceleration times (as in the Hwa-Bjorken scenario).
If the acceleration of the point charge is short,
a bell-shaped profile appears, proportional to $1/\cosh^4\eta$.


With this paper we would like to call attention to the possibility
that neither a collective flow nor a thermal state has to be
necessarily assumed in order to produce particle spectra
resembling such behavior. Experimentally a decision can probably be
made by hunting for the occurence of an interference pattern
in the photon transverse momentum distribution, which - according to
simple calculations presented here - emerge for certain deceleration scenarios
at certain $k_{\perp}$-s. We use units in which $k_B=\hbar=c=1$.

\section{Radiation from accelerating point charge}

It is well known 
\cite{Itzyckson+Zuber,LandauFiz} 
that the solution of the Maxwell equations in
radiation (Lorenz) gauge, or equivalently the use of the Feynman propagator
delivers a radiation spectrum equivalent to the following photon number distribution
\be
{d^3 N} = \frac{1}{2k_0} \frac{d^3k}{(2\pi)^3} \sum \left|\epsilon^{(a)}\cdot J(k) \right|^2
\ee{TRADI_RADI}
with  $J(k)$ being the Fourier transform of the four-current of the charged source of
the radiation and the summation going over two transverse polarization states. The four-momentum
$k_i$ is taken on mass shell, i.e. $k\cdot k =k_ik^i=0$.
By considering a point charge $q$ moving along a foregiven trajectory $x_i(\tau)$,
parametrized with the proper time $\tau$, the source current density is usually taken
as a Dirac-delta constraint on that trajectory. This results in
\be
J^i(k) = q \int e^{ik\cdot x(\tau)} \, u^i(\tau) \, d\tau,
\ee{TRADI_CURR}
with $u^i(\tau)=dx^i(\tau)/d\tau $ being the normalized tangential to the worldline,
the four-velocity of the moving point. In this way
the Fourier transform is integrated over the worldline history of the point charge.

However this often used formula for obtaining the irradiated photon spectra
is valid only when the integration limits are minus and plus infinity. For a finite proper
time history one should be more careful \cite{Jackson}. 
The above prescription namely would give
a non-vanishing contribution also for a charge moving with constant velocity, although
this should not radiate. The resolution of this problem lies in considering the
partial integration formula,
\be
\int e^{ik \cdot x} \td{}{\tau} \left( \frac{u^i}{k\cdot u} \right) \, d\tau = 
\left. e^{ik \cdot x}  \frac{u^i}{k\cdot u} \right|_{\tau_1}^{\tau_2} -  
\int \td{}{\tau} \left( e^{ik \cdot x}\right)  \frac{u^i}{k\cdot u} \, d\tau  
\ee{PARTINT}
Executing the derivation of the plane-wave factor cancels the denominator and one obtains
\be
\int e^{ik \cdot x} \td{}{\tau} \left( \frac{u^i}{k\cdot u} \right) \, d\tau = 
\left. e^{ik \cdot x}  \frac{u^i}{k\cdot u} \right|_{\tau_1}^{\tau_2} -  
i \, \int  \left( e^{ik \cdot x}\right)  {u^i} \, d\tau  
\ee{PART_INT_USEFUL}
Since the left hand side above vanishes for non-accelerating motion,
we use this instead of eq.(\ref{TRADI_CURR}) to calculate the number of radiated photons.
With other words dropping the contributions at
the initial and final time instants we have in mind that the charge was moving
with the respective constant velocities before and after the finite acceleration
(or deceleration) period. These considerations lead us to the use of
the following projected Fourier transform
\be
\epsilon \cdot J(k) = q \int_{\tau_1}^{\tau_2}\!e^{ik\cdot x(\tau)}  \,
	\td{}{\tau} \left(\frac{\epsilon\cdot u}{k \cdot u}\right) \, d\tau.
\ee{TRANSVERSE_FOURIER}
By inspecting
\be
  \td{}{\tau} \left(\frac{\epsilon\cdot u}{k \cdot u}\right) =
  \frac{(\epsilon\cdot a)(k\cdot u)-(\epsilon\cdot u)(k\cdot a)}{(k\cdot u)^2},
\ee{DERIV_EPSU}
it is clear that only the accelerating charges contribute to the radiation.
Here $a^i=du^i/d\tau$ is the acceleration four-vector.

It is important to realize that this result on the spectrum of photons 
is also valid in the quantum theory of photons. Then the probability to create
an $n$-photon state from a zero-photon state with given four-momentum $k$ is
Poisson distributed, with the mean value being the classical result.


Summarizing the result of the above considerations the Lorentz-invariant photon spectrum
is given as
\be
\phott = \frac{2\alpha_{EM}}{\pi} \, \sum \left|{\cal A} \right|^2
\ee{PHOTSPECTR}
with
\be
{\cal A} = \int_{\tau_1}^{\tau_2}\limits\!e^{ik\cdot x(\tau)} \, 
   \td{}{\tau} \left( \frac{\epsilon\cdot u}{k\cdot u}\right) \, d\tau.
\ee{AMPLITUDE}
Considering straight line motion for the point charge with an acceleration parallel to the velocity,
but with finite initial and final velocities, we parametrize the essential four vectors as follows.
The photon four-momentum on mass shell is given by
\be
 k_i = k_{\perp} \left(\cosh\eta,\sinh\eta,\cos\psi,\sin\psi \right).
\ee{PHOTON4}
We take two orthogonal spacelike polarization vectors:
\be
\epsilon^{(1)}_i  =  \left( \sinh\eta, \cosh\eta, 0, 0 \right), 
\qquad
\epsilon^{(2)}_i  =  \left( 0, 0, -\sin\psi, \cos\psi \right) 
\ee{POLAR4}
The four-velocity of the source points to the first direction:
\be
 u_i = \left( \cosh\xi,\sinh\xi, 0, 0 \right)
\ee{VELOC4}
The four-acceleration is given by its $\tau$-derivative:
\be
a_i = \td{u_i}{\tau} = \left( \sinh\xi,\cosh\xi, 0, 0 \right) \, \td{\xi}{\tau}.
\ee{ACCEL4}
In this paper we shall consider only constant proper decelerations, $d\xi/d\tau=-g$,
independent of $\tau$ and plot results for $g=1$. 
This simplifies a lot. However, since we calulate our spectra for
arbitrary proper time intervals, any acceleration profile could in principle be reconstructed
numerically based on the present results.

The only non-vanishing combination occurring in the formula (\ref{DERIV_EPSU}) 
for the photon spectrum is given as
\be
(\epsilon^{(1)}\cdot a)(k\cdot u) - (\epsilon^{(1)}\cdot u)(k\cdot a) = g k_{\perp}.
\ee{RELEVANT_TERM}
The amplitude is finally given as
\be
{\cal A} =  \frac{1}{k_{\perp}} \, 
  \int_{\tau_1}^{\tau_2}\limits e^{ik\cdot x(\tau)} \, \frac{gd\tau}{\cosh^2(\xi-\eta)}.
\ee{FINAL_AMPLI}
Inspecting this result, it becomes transparent that the most suited integration
variable is a (Lorentz transformed) velocity, $v=\tanh(\xi-\eta)$.
Using this the amplitude becomes
\be
{\cal A} = \frac{e^{i\varphi_0}}{k_{\perp}} \, \int_{v_1}^{v_2}\limits e^{i\ell k_{\perp}\gamma v} dv,
\ee{AMPLI_BECOMES}
with $\ell=1/g$ and a $\varphi_0=k\cdot x(0)$ phase corresponding to the initial position.
The limits are to be taken at $v_i=\tanh(\xi_i-\eta)$.


The calculation runs between the proper time points $\tau_1$ and $\tau_2$, with variable
reference rapidity $\xi_0$, defining $\xi_1=\xi_0+g\tau_1$ and $\xi_2=\xi_0+g\tau_2$. 
Since the photon-rapidity dependence enters into
the calculation under an integral in the form of $\xi-\eta$ only, the resulting photon spectrum
is a function of it via the starting and final rapidity differences:
$\xi_1-\eta$ and $\xi_2-\eta$. The calculation at changing $\xi_0$ and fixed $\eta=0$ and 
$\tau_1+\tau_2=0$ therefore completely reveals the $\eta$-dependence 
when the photon yield is plotted against the variable
\be
  \xi_{{\rm mid}} -\eta = \xi_0 + g\frac{\tau_1+\tau_2}{2} - \eta.
\ee{MID_PHOTON_RAPID}

Let us first investigate some analytically handy cases.

First we note that for small transverse momenta of the photon,
$\ell k_{\perp} \ll 1$, the $k_{\perp}^2$ times photon yield approaches
a constant value. This value depends on the length of integration,
on the rapidity gap, $\xi_2-\xi_1$, during which the deceleration of the source is active.

Considering that  
\be
k_{\perp}^2 \, \phott = 2\alpha 
\left|\int_{\xi_1-\eta}^{\xi_2-\eta} \! e^{i\ell k_{\perp} \sinh \xi} \frac{d\xi}{\cosh^2\xi} \right|^2,
\ee{kT2YIELD}
the small $k_{\perp}$ approximation is an analytically calculable integral.
Rewriting in terms of a velocity integral,
\be
k_{\perp}^2 \, \phott = 2\alpha 
\left|\int_{v_1}^{v_2} \! e^{i\ell k_{\perp} \gamma v} {dv} \right|^2.
\ee{VEL_kT2YIELD}
Its non-relativistic approximation is obtained by setting $\gamma=1$:
\be
k_{\perp}^2 \, \phott =  
\frac{8\alpha}{\ell^2 k_{\perp}^2} \, \sin^2 \left( \ell k_{\perp}\frac{v_2-v_1}{2}\right) . 
\ee{NONREL}
This result incorporates non-trivial interference effects. 
It also shows that the Lorentz invariant photon spectrum is always smaller than 
an estimate which is proportional to the inverse 4-th power of the photon transverse momentum,
\be
\phott \le \frac{8\alpha}{\ell^2 k_{\perp}^4}. 
\ee{RUTHERFORD}

The generic infrared result for $k_{\perp}=0$ at arbitrary initial and final velocities
is given by the velocity difference squared:
\be
\lim_{k_{\perp}\rightarrow 0}\limits \: k_{\perp}^2 \, \phott = 2\alpha |v_2-v_1|^2.
\ee{VEL_DIFF_YIELD}
Expressing this in terms of the rapidity variables of the source at the
beginning and at the end of integration, $v_i=\tanh(\xi_i-\eta)$, one notes
that
\be
\Delta v := \frac{1}{2} (v_2-v_1) = \frac{1}{2} \left(\tanh(\xi_2-\eta)-\tanh(\xi_1-\eta) \right)
\ee{VEL_DIFF_CH}
can be written in terms of hyperbolic sine and cosine functions as
\be
\Delta v = \frac{\sinh(\xi_2-\eta)\cosh(\xi_1-\eta)-\sinh(\xi_1-\eta)\cosh(\xi_2-\eta)}{2\cosh(\xi_1-\eta)\cosh(\xi_2-\eta)}.
\ee{DELTA_V_AS_SH}
Applying now well known relations for the hyperbolic functions we arrive at
\be
\Delta v = \frac{\sinh\frac{\xi_2-\xi_1}{2} 
  \, \cosh\frac{\xi_2-\xi_1}{2}}{\cosh^2(\xi_{{\rm mid}}-\eta)+\cosh^2\frac{\xi_2-\xi_1}{2}-1}.
\ee{DELTAV_FINAL}
In this formula the $\eta$-dependence is well separated in the variable $\xi_{{\rm mid}}-\eta$.
Therefore the rapidity distribution at very small $k_{\perp}$ becomes
\be
\lim_{k_{\perp}\rightarrow 0}\limits \: 
k_{\perp}^2 \, \phott = 8\alpha 
\left( \frac{\sinh\frac{\xi_2-\xi_1}{2} 
\, \cosh\frac{\xi_2-\xi_1}{2}}{\cosh^2(\eta-\xi_{{\rm mid}})+\cosh^2\frac{\xi_2-\xi_1}{2}-1} \right)^2.
\ee{IR_RAPDI}

Two limiting cases of this formula can be of interest. For small diferences between the
initial and final rapidities of the charge one obtains a bell-shaped  form
\be
\lim_{k_{\perp}\rightarrow 0}\limits \: 
k_{\perp}^2 \, \phott = {2\alpha} \:  \frac{(\xi_2-\xi_1)^2}{\cosh^4(\eta-\xi_{{\rm mid}})}, 
\ee{IR_LANDAU}
resembling the features of the rapidity distribution obtained by using Landau's
hydrodynamical model \cite{LANDAU:HYDRO}.
On the other hand, for very large differences between the final and initial rapdities of the
radiation source, this quantity approaches a constant. This is compatible to the Unruh scenario
discussed previously \cite{Biro+Schram+Gyulassy}:
\be
\lim_{k_{\perp}\rightarrow 0}\limits \: 
k_{\perp}^2 \, \phott \propto 8\alpha  
  \frac{1}{\left(1+\varepsilon \sinh^2(\eta-\xi_{{\rm mid}}) \right)^2}, 
\ee{IR_BJORKEN}
with $\varepsilon=\exp(-|\xi_2-\xi_1|)$.
This result represents an elongated plateau in the rapidity distribution, reminding
to the Hwa-Bjorken hydrodynamical scenario \cite{BJORKEN,HWA}.



\section{Results on differential rapidity distributions}

\begin{figure}
\begin{center}
        \includegraphics[width=0.44\textwidth,angle=-90]{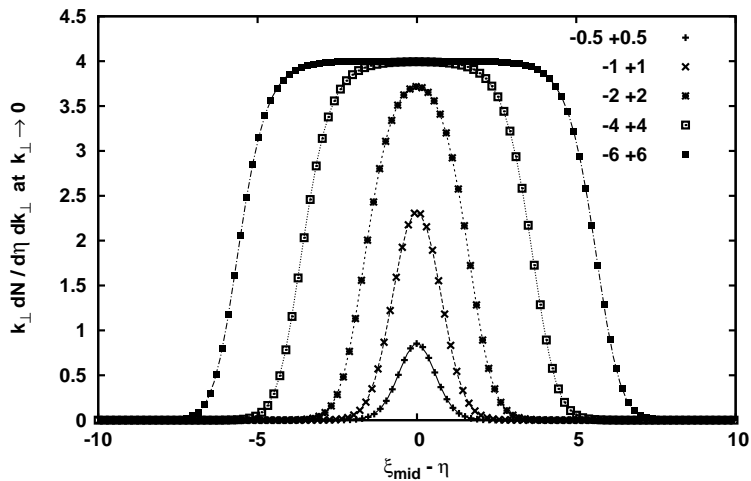}
\end{center}
\caption{ \label{FIG:RAPALL}
	$k_{\perp}^2$ times the invariant photon yield at $k_{\perp}=0.01$ 
	as a function of the rapidity $\xi_{{\rm mid}}-\eta$. 
	This numerical value we use as an approximation for the infrared limit.
	The different curves belong
	to varying proper time durations of the constant acceleration ($g=1$) 
	according to the legend (denoting $\tau_1$ and $\tau_2$ values). 
}
\end{figure}

Figure \ref{FIG:RAPALL} features the photon yield at low $k_{\perp}$ 
as the rapidity distribution is a function of $\xi_{{\rm mid}}$ (cf. eq.\ref{MID_PHOTON_RAPID}).
Here the short interval decelaration causes a smaller yield,
with a bell-shape, familiar from Landau's hydrodynamic scenario.
Longer term constant deceleration let a plateau  develop
in this curves, reminding us to the Hwa-Bjorken hydrodynamical scenario.
The continous lines follow the exact formula (\ref{IR_RAPDI}).

\begin{figure}
\begin{center}
        \includegraphics[width=0.30\textwidth,angle=-90]{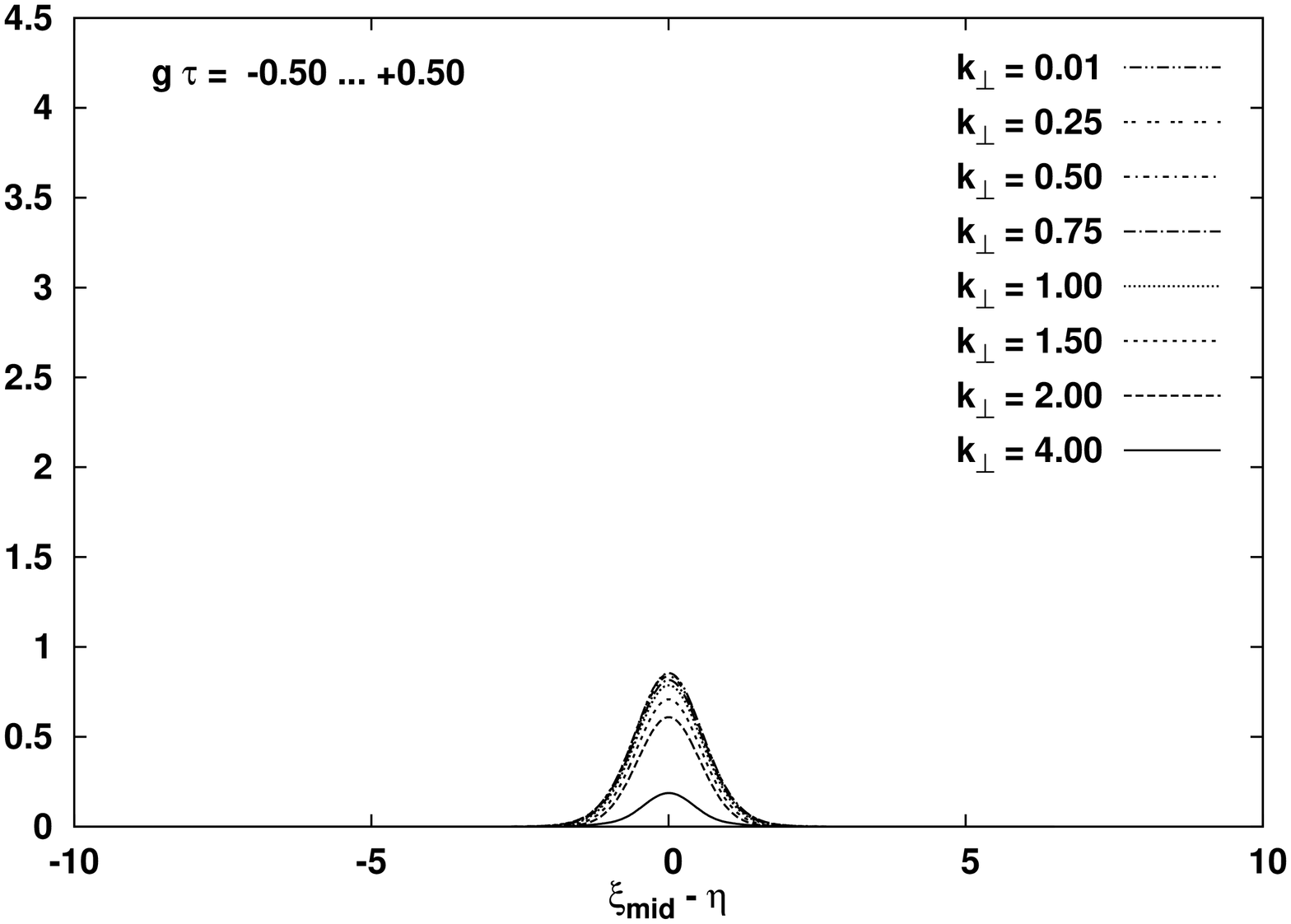}
	\includegraphics[width=0.30\textwidth,angle=-90]{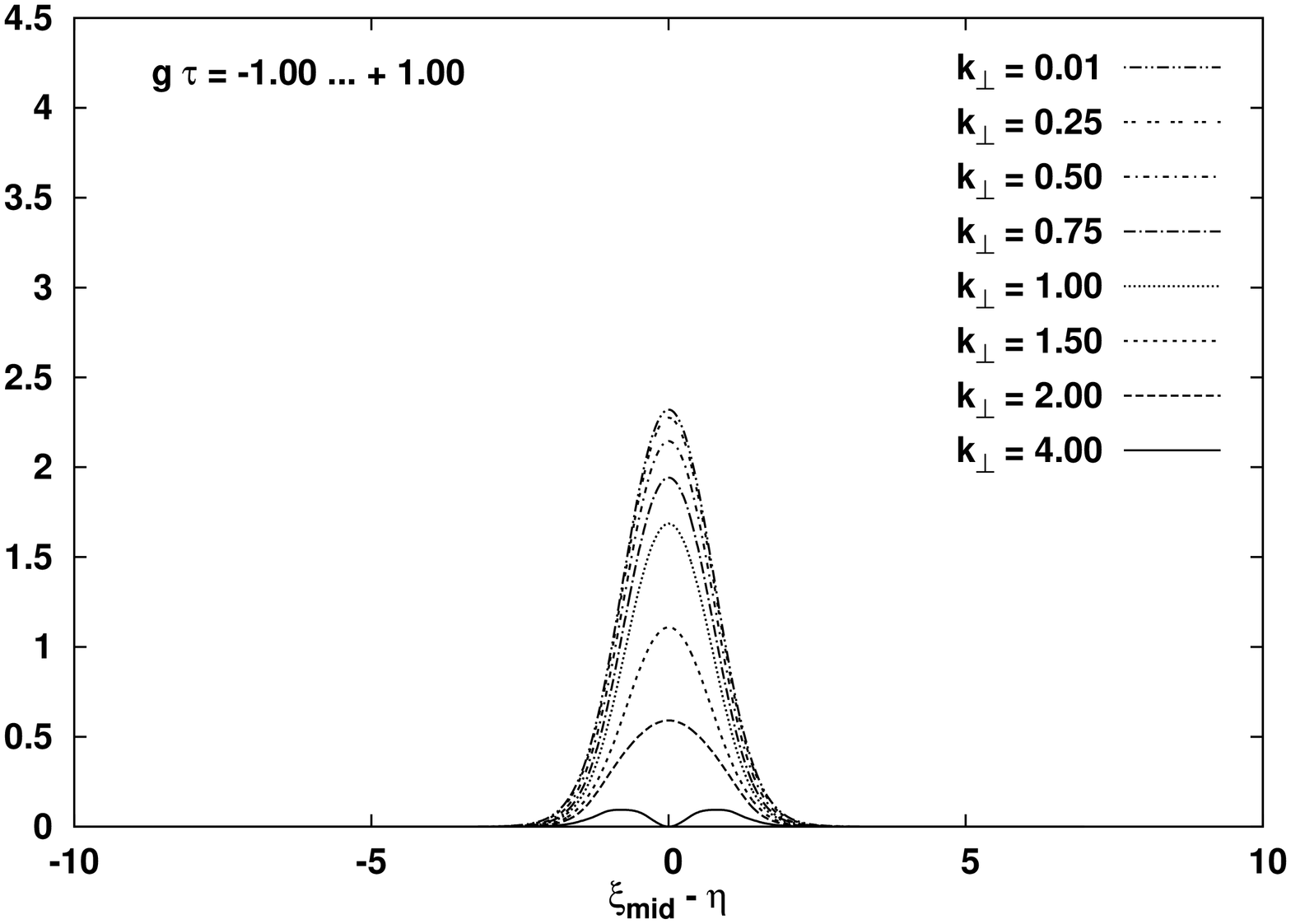}
        
	\includegraphics[width=0.30\textwidth,angle=-90]{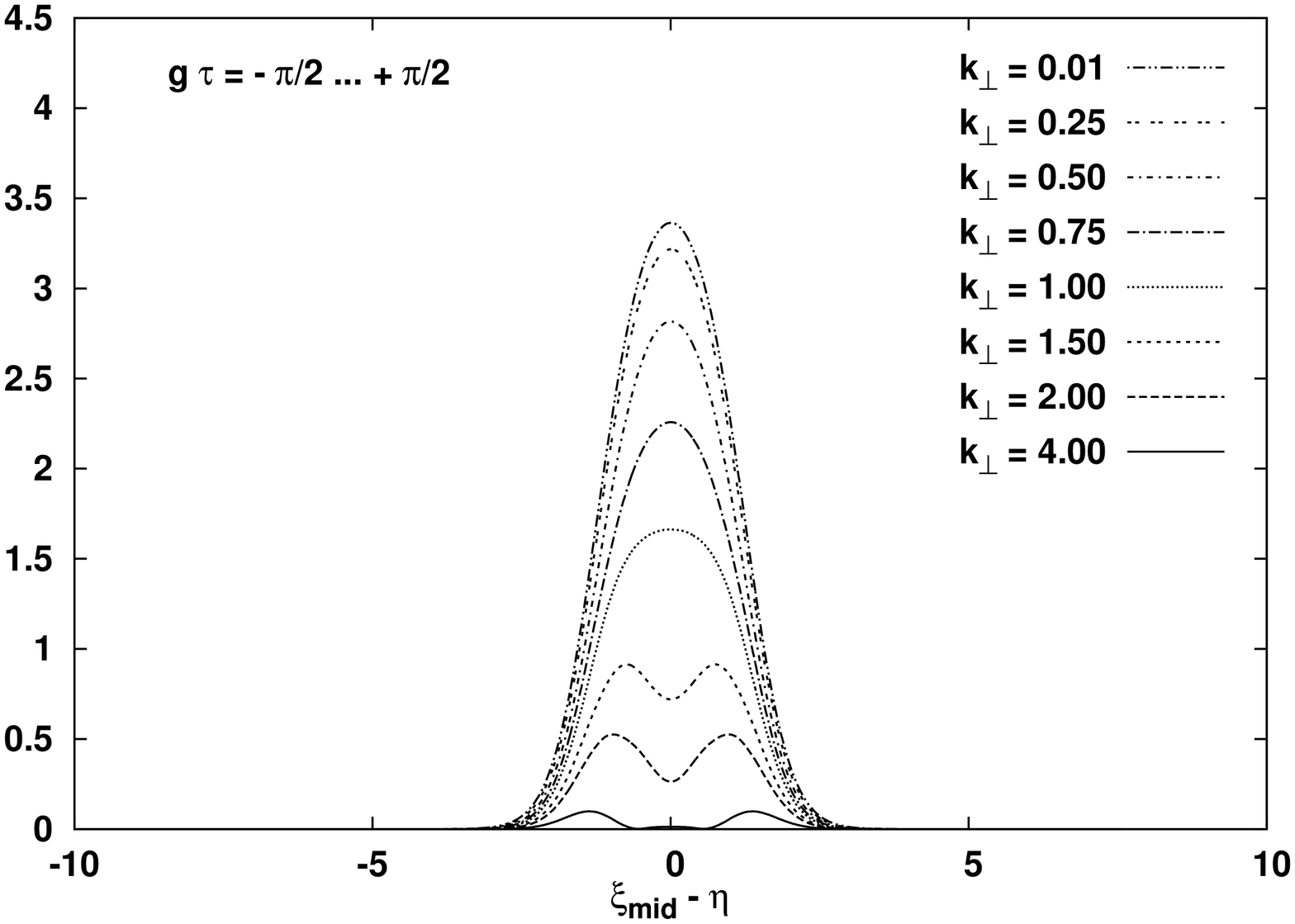}
	\includegraphics[width=0.30\textwidth,angle=-90]{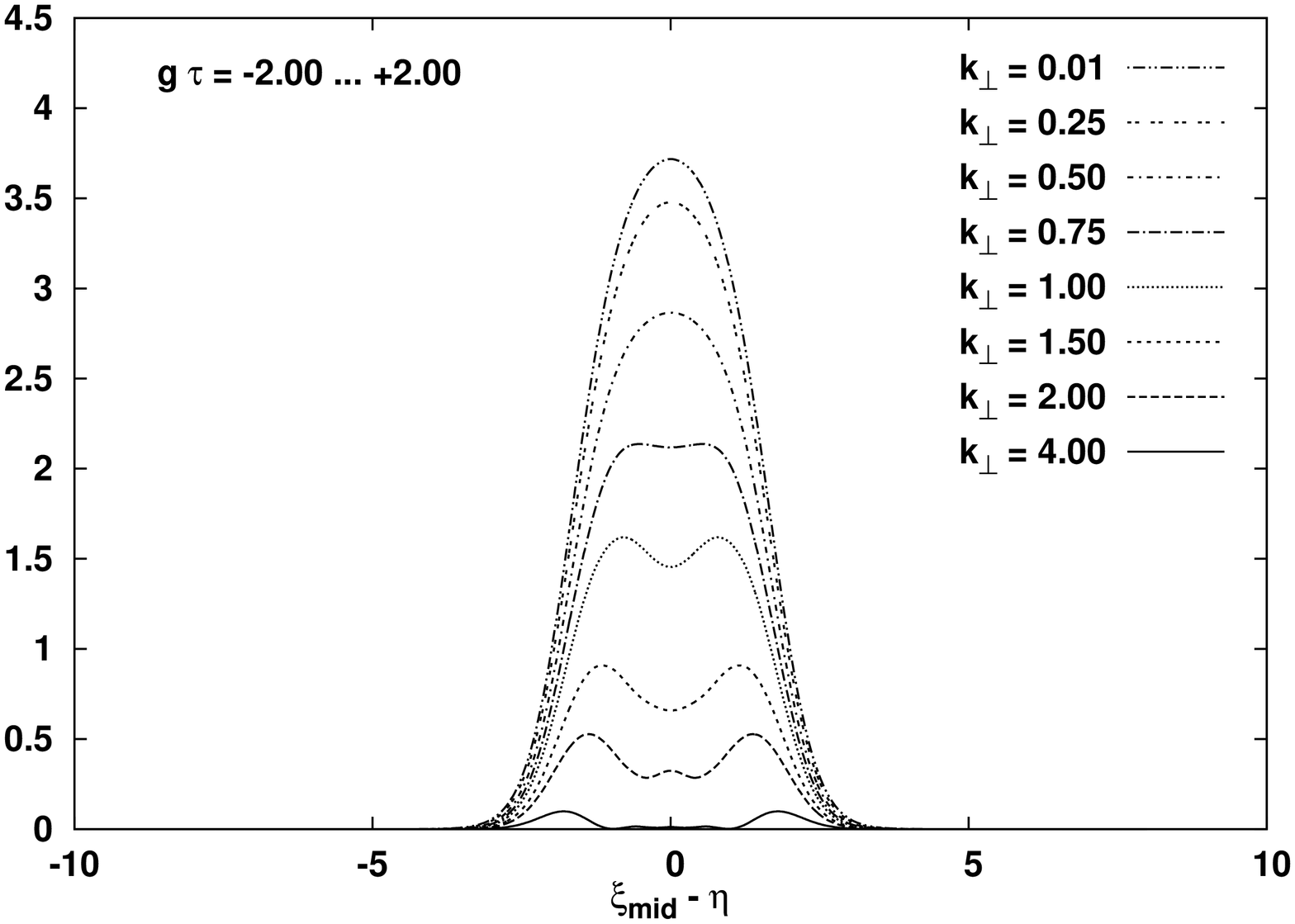}
\end{center}
\caption{ \label{FIG:RAP6_1}
	Photon yield multiplied with $k_{\perp}^2$ -- rapidity distributions for several
	short deceleration times $g\tau = 0.5, 1, \pi/2, 2$. 
	The different curves belong to
	different $k_{\perp}$ values at $g=1$ according to the legend.
}
\end{figure}

\begin{figure}
\begin{center}
        \includegraphics[width=0.33\textwidth,angle=-90]{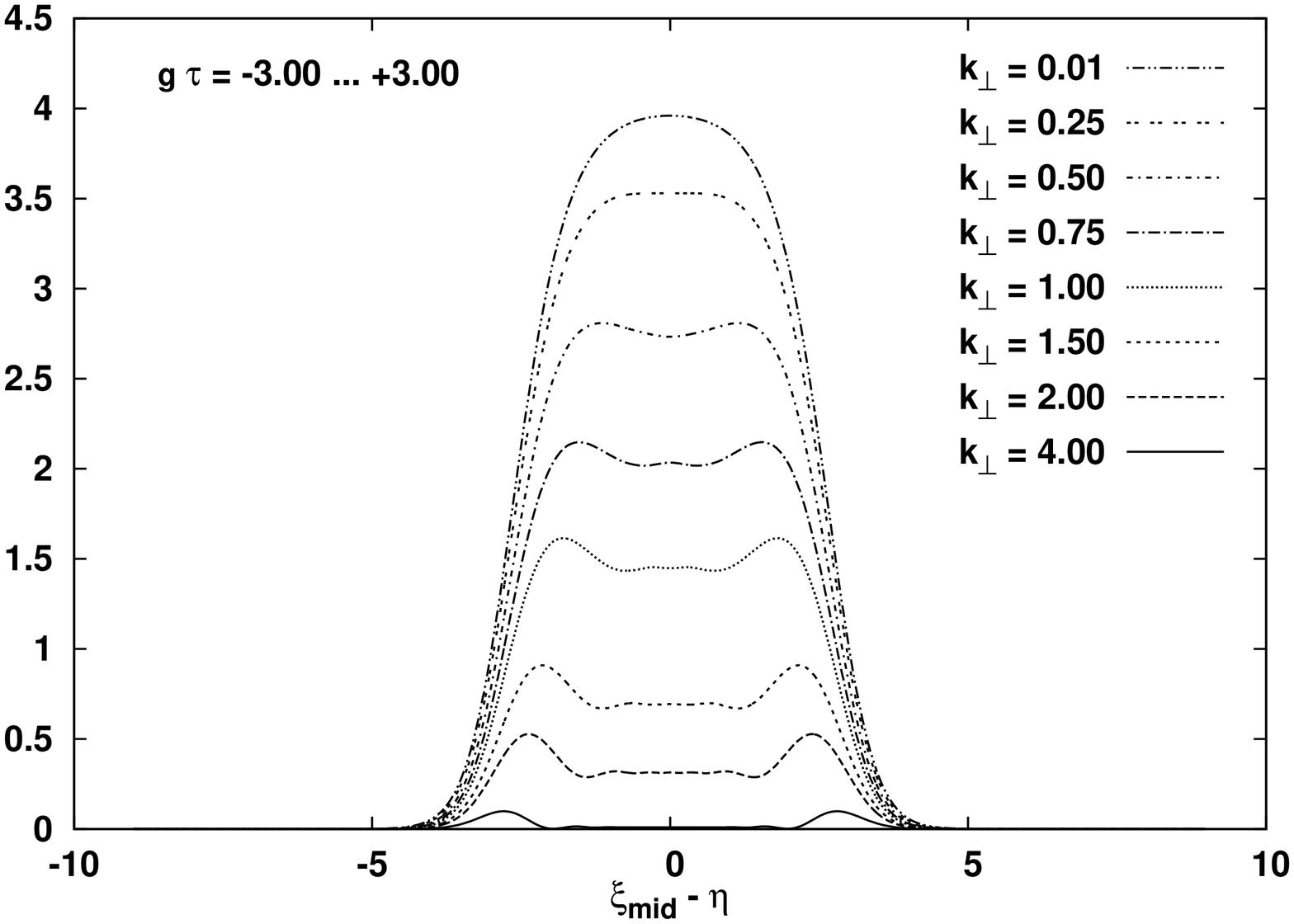}
	\includegraphics[width=0.33\textwidth,angle=-90]{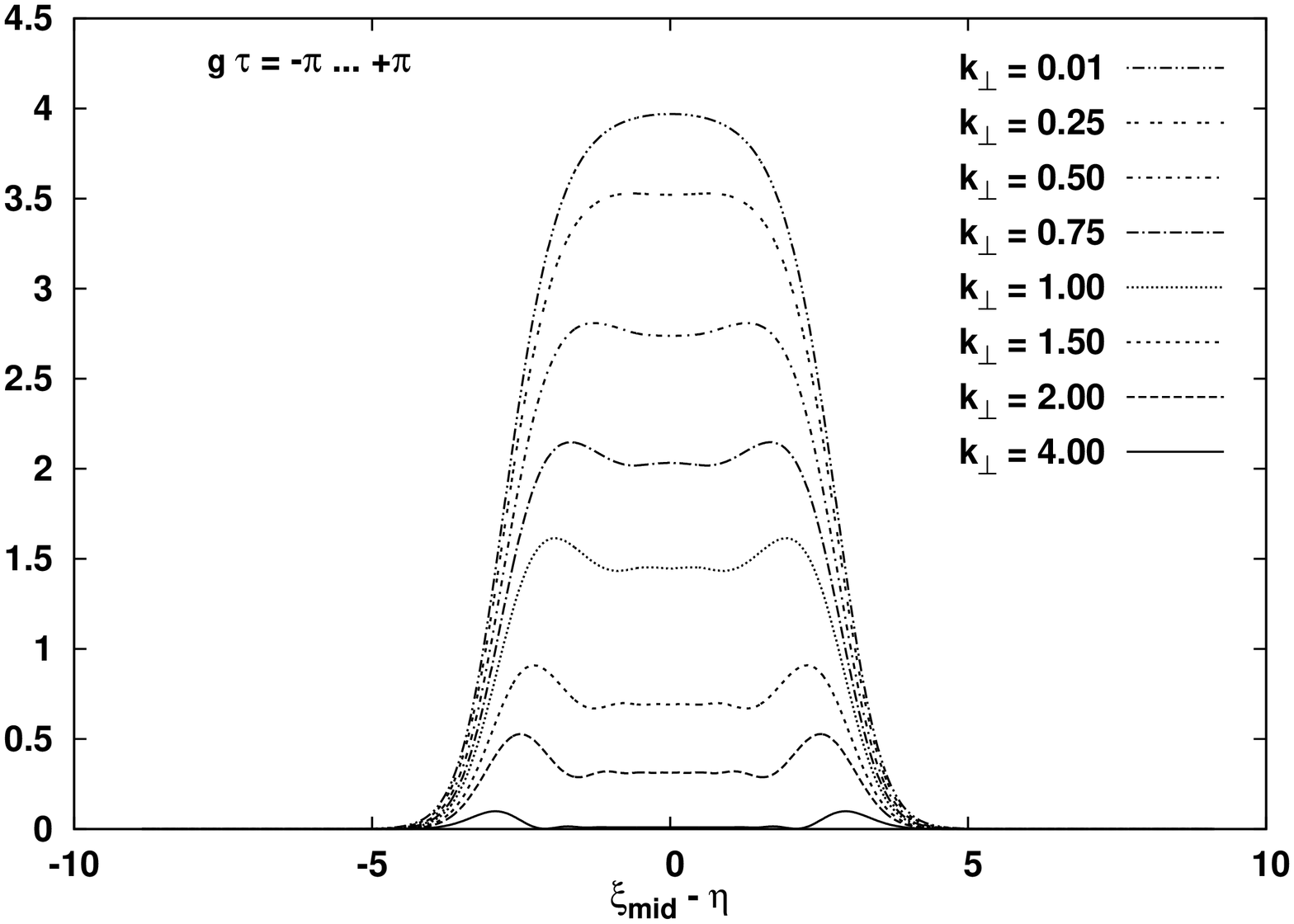}
        
	\includegraphics[width=0.33\textwidth,angle=-90]{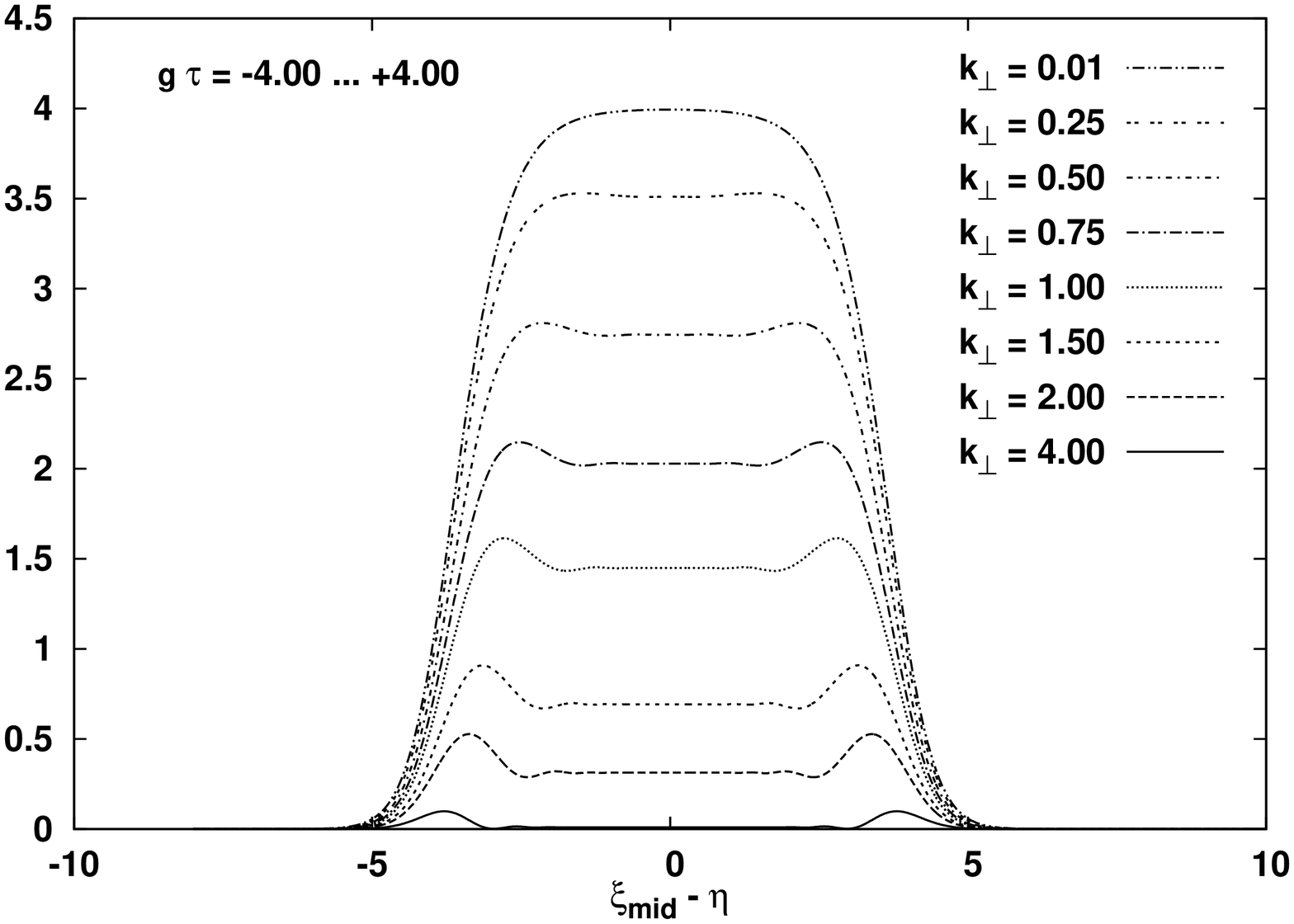}
	\includegraphics[width=0.33\textwidth,angle=-90]{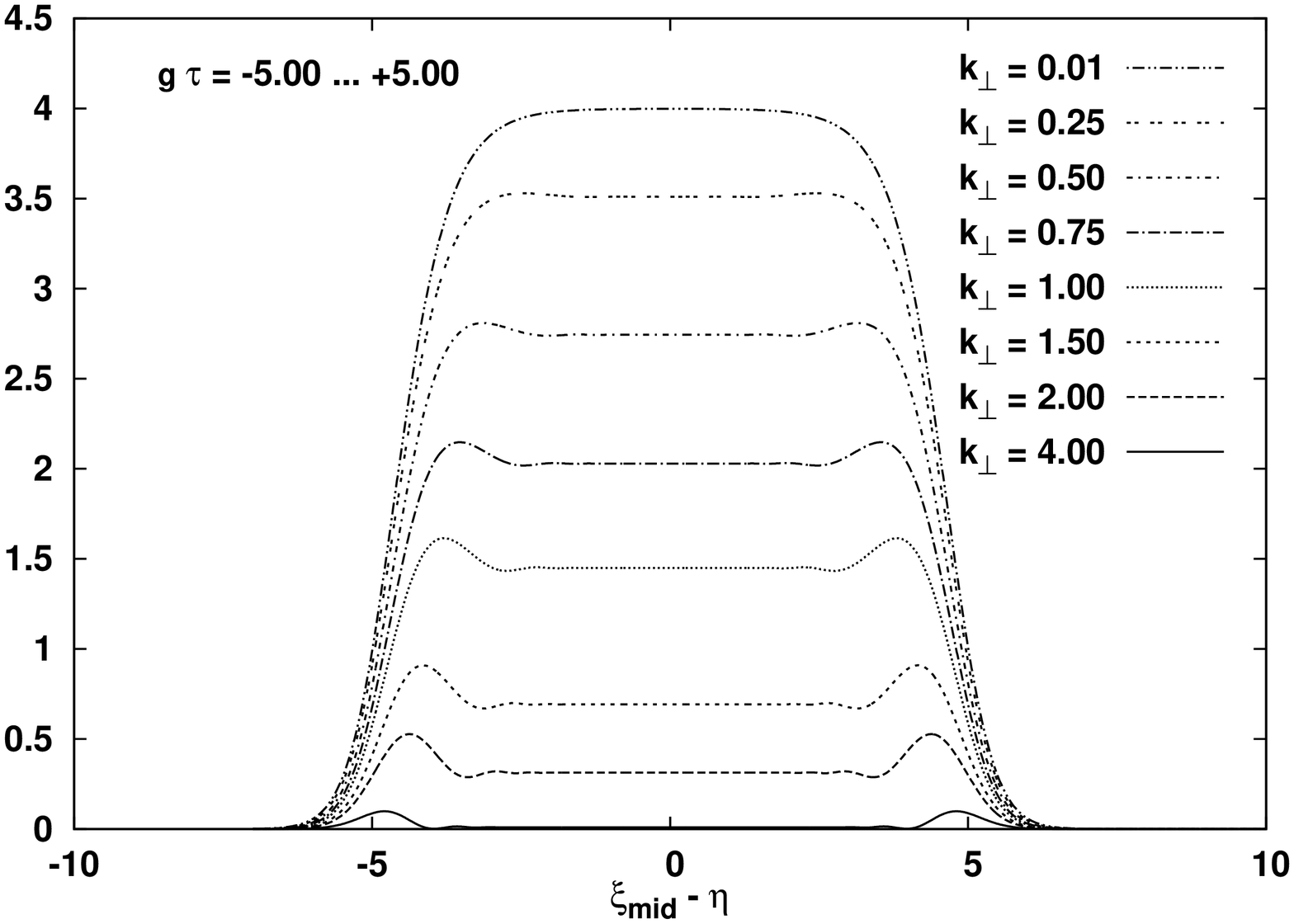}
\end{center}
\caption{ \label{FIG:RAP6_5}
	Photon yield multiplied with $k_{\perp}^2$ -- rapidity distributions 
	for longer deceleration times ($g\tau = 3, \pi, 4, 5$). 
	The different curves belong to
	different $k_{\perp}$ values with $g=1$ according to the legend.
}
\end{figure}

We are also interested in the photon double differential yield (multiplied by
$k_{\perp}^2$ for the sake of de-emphasizing the infrared divergence) in the
classical radiation picture. Two examples are shown in the following figures:
one for short time constant acceleration from proper time $-\tau$ to $+\tau$ 
(Fig.\ref{FIG:RAP6_1}), $g\tau=0.5,1.0,\pi/2,2.0$, 
and one for long acceleration (Fig.\ref{FIG:RAP6_5}), $g\tau=3.0,\pi,4.0,5.0$.
The transition between plateau and non-plateau behavior can be observed at all
$k_{\perp}$ values leading to considerable yields. Moreover one realizes that
around $\ell k_{\perp} \approx 1/2$ with $\ell = 1/g$ an interference pattern starts to develop
at the edge of the rapidity plateau. This is a remarkable feature, and probably
distinguishes a pseudothermal mechanism, like discussed here, from a real thermal
equilibrium spectrum looking alike a black body radiation (having just the plateau,
an $\eta$-independent yield for $\tau_1\rightarrow-\infty, \tau_2\rightarrow+\infty$).

The big jump between ''short'' and ''long'' acceleration behavior
(between kicks and Unruh-type scenarios) seems to occur around
$g\tau \approx \pi$, when the phase under the integral (\ref{AMPLI_BECOMES})
takes a whole period of interference into account.

Our results discussed so far clearly show that the calculated photon
rapidity spectra are qualitatively similar to those obtained in
hydrodynamical models.
A recent example of this more common approach for calculating rapidity spectra
for massive particles is shown by Jiang \cite{Jiang}. The idea is to use a solution of 
hydrodynamical equations for a fluid medium for obtaining the entropy density, 
and from that the rapidity distributions 
for nucleus-nucleus collisions, which supposed to be proportional to the entropy
density. This consideration once leads to the Hwa-Bjorken scenario, if
the original rapidity is set equal to the space rapidity (boost invariant flow secnario), 
otherwise it resembles results from the boost non-invariant case. 
In the aftermath of such calculations the solution is applied to determine
the rapidity dependent entropy density, as being proportional to the rapidity distributions.


\begin{figure}
\begin{center}
        \includegraphics[width=0.40\textwidth,angle=-90]{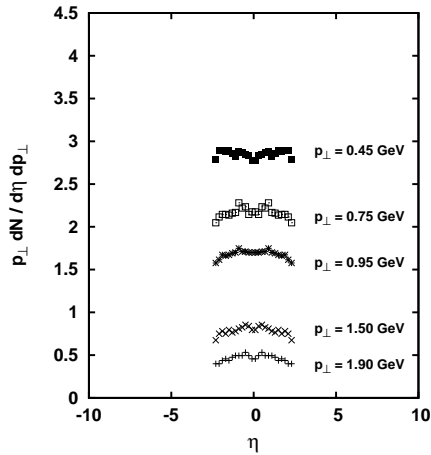}
\end{center}
\caption{ \label{FIG:CMS}
	Experimental hadron inclusive invariant yields multiplied with $p_{\perp}^2$ vs
	rapidity distributions at several $p_{\perp}$ values as 
	measured in pp collisons by the CMS at $\sqrt{s_{NN}}= 7$ TeV. Data for the plot
	are taken from Ref.\cite{CMSpp}.
}
\end{figure}

\begin{figure}[H]
\begin{center}
        \includegraphics[width=0.33\textwidth,angle=-90]{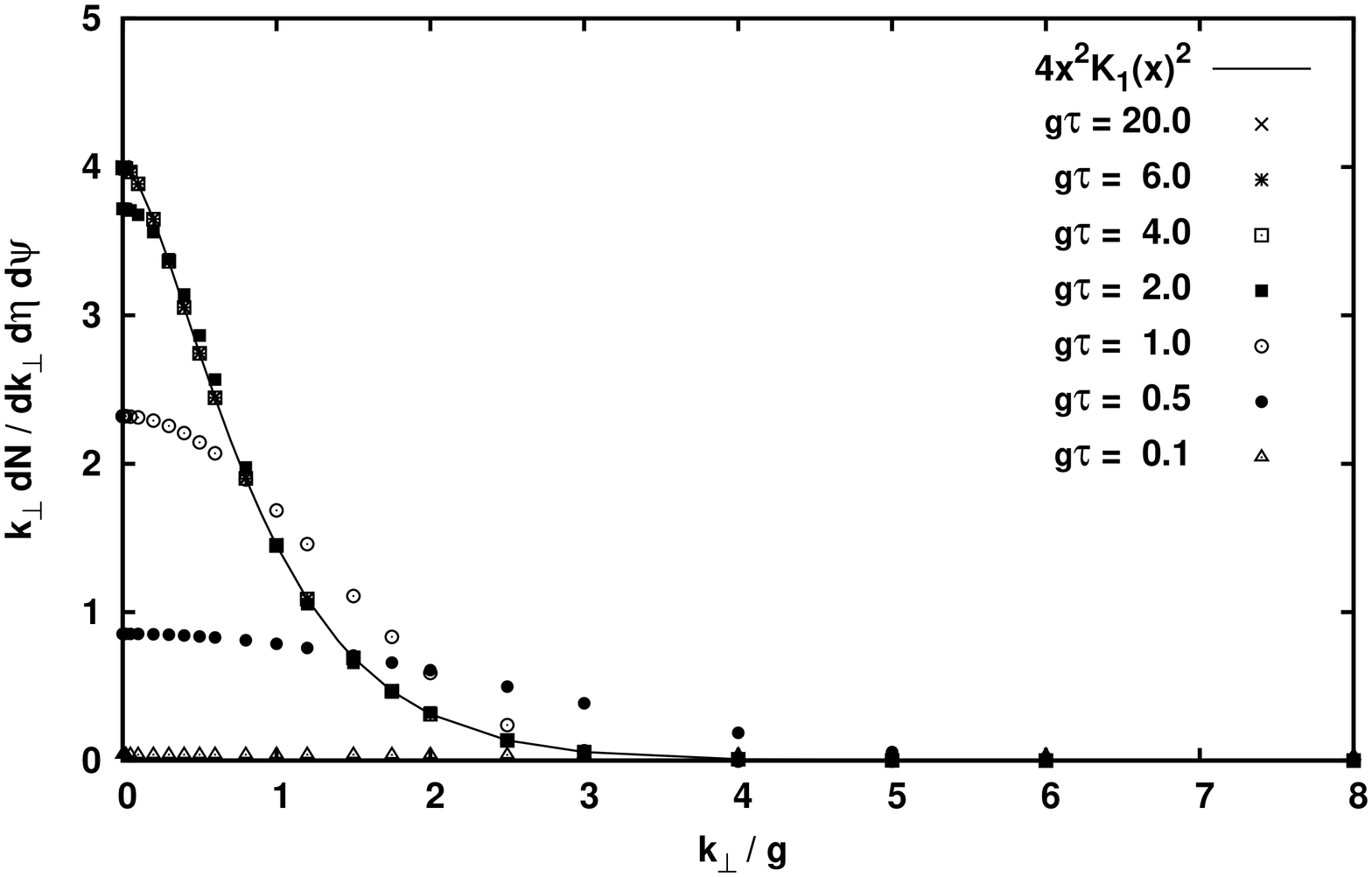}
        \includegraphics[width=0.33\textwidth,angle=-90]{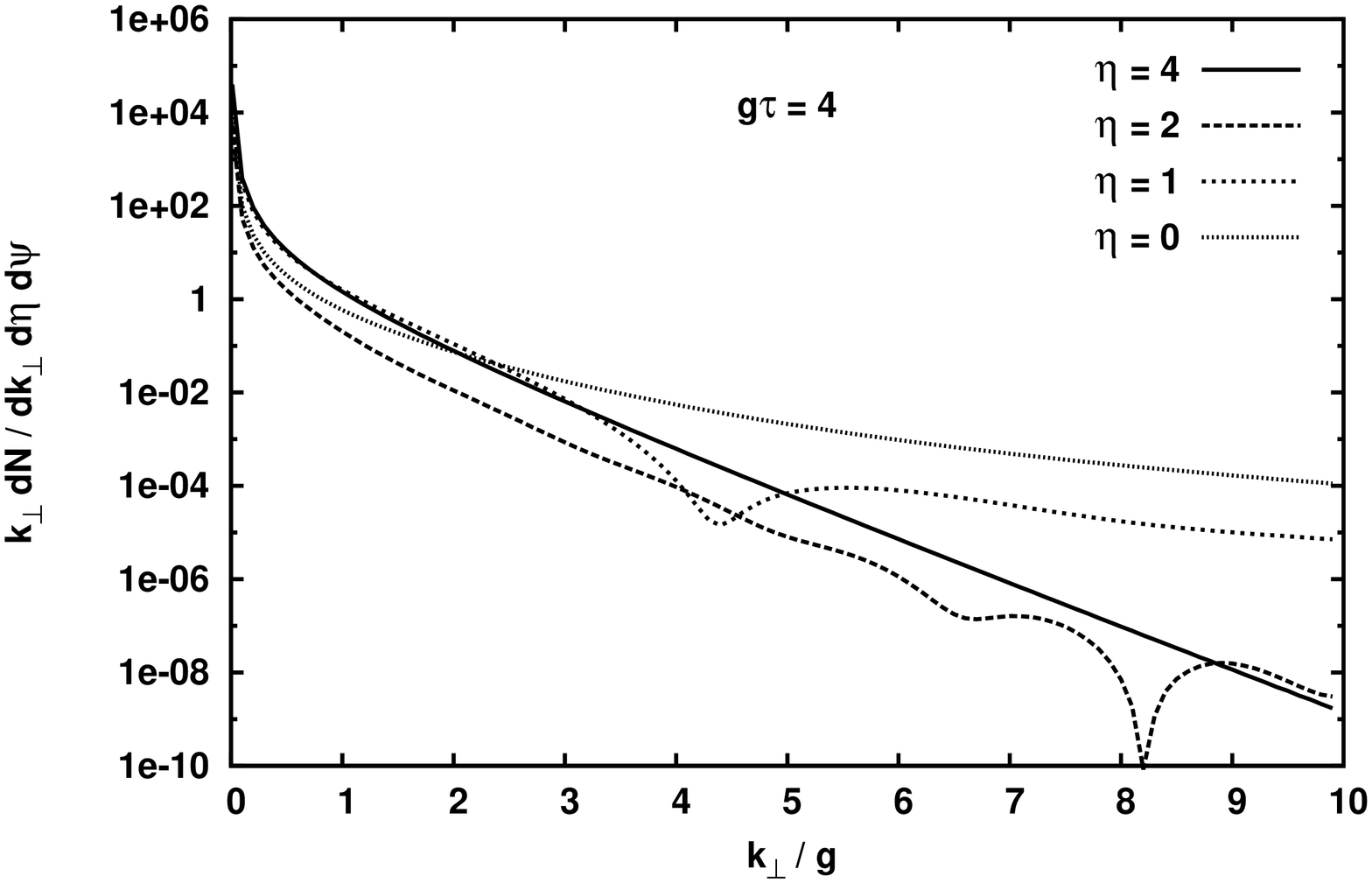}
\end{center}
\caption{ \label{FIG:KTCOULOMB}
	Left:
	$k_{\perp}^2$ times the invariant photon yield as a function of
	the transverse momentum $k_{\perp}$. The different points belong
	to varying integration proper rapdities from $-g\tau$ to $g\tau$
	according to the legend ($g=1$). The analytic result
	published Ref.\cite{Biro+Schram+Gyulassy} is represented by
	a continous line.
	Right:
	logarithmic plot of the invariant photon yield as a function of
	$k_{\perp}$ for different $\eta-\xi_{{\rm mid}}$ arguments
	denoted briefly as $\eta$ in the legend. 
}
\end{figure}

Fig.\ref{FIG:CMS} displays results from CERN LHC, measured by the
CMS collaboration at $7$ TeV for $pp$ to all hadrons spectra \cite{CMSpp}. 
Of course hadrons are  - unlike photons -- massive, but at high enough transverse
momenta, $p_T$, this should not matter much. Also polarization factors
may behave differently. The experimental data show plateau-like behavior
and a monotonic decrease of yields from $p_T=0.45$ GeV upwards.
Enhancement at edge rapidities might be sensed at low transverse momentum
in the differential rapidity distributions, usually tagged to ''transparency''.
It is however not obvious why would be the transparency larger at low $p_T$
than at high $p_T$, since soft cross sections tend to be larger than the
hard ones. 
Fig.\ref{FIG:KTCOULOMB} plots the $k_{\perp}$-dependence for various
acceleration times.
On the left side the scaled invariant photon yield is seen
for different integration intervals for the moving source
from $-g\tau$ to $g\tau$ according to the legend ($g=1$). 
The analytic result published in Ref.\cite{Biro+Schram+Gyulassy} is represented by
a continous line, it is approached well already for $g\tau=4$.
On the right side we present a logarithmic plot of the invariant photon yield as a function of
$k_{\perp}$ for different $\eta-\xi_{{\rm mid}}$ arguments
denoted briefly as $\eta$ in the legend. 
Here some interference pattern can be observed at higher transverse momenta.

Throughout this paper we used the value $g=1$ so only the shapes of the spectra
shown are relevant for discussion. The proper time values for $\tau$
are $g\tau/c$ values in the general case and any $k_{\perp}$ value indicated above
transforms to $\ell k_{\perp}=c^2 k_{\perp}/g$. To set the basic scale in physical
units the typical stopping length, $\ell=c^2/g$, is estimated to be less or in the order
of magnitude of the target size. Fitting gamma spectra at RHIC
we obtained earlier an equivalent Unruh temperature of $135$ MeV,
corresponding to a chracteristic deceleration length of $\ell = \hbar c/(2\pi T_U) \approx 0.22$ fm.
The exponential slope parameter (fitted to the experimental data)
in that case was $T=\pi T_U \approx 424$ MeV.



\section*{Summary}

Based on the above calculations we conclude that from experiencing 
flat or bell-shaped rapidity distributions
of secondary light particles, in particular photons, one should not infer
the presence of a flowing source medium. This caution may be proper also for
considering massive particle spectra if the observed transverse momenta are
essentialy larger than the rest mass. 

Experimental data show distributions for several
particles from nucleus-nucleus collisions with features of the spectra similar to those 
seen in Fig.\ref{FIG:RAP6_5} (higher $k_{\perp}$-s).
The difference between statistical scenarios with collectively flowing sources
and near-classical field theory calculations can, however, in principle be
experimentally investigated: at certain rapidities the photon transverse spectra will show
interference patterns with characteristic dips in the second case.

\section*{Acknowledgment}

This work has been supported by the Hungarian National Research Fund (OTKA K104260)
and by the Helmholtz International Center for FAIR within the framework of the
LOEWE program launched by the State of Hesse.


\begin{thebibliography}{xxxx}

\bibitem{HAGEDORN}  R.~Hagedorn, Suppl. Nuovo Cim. 3 (1965) 147 

\bibitem{Hagedorn2} R.~Hagedorn: Nucl. Phys. B 24 (1970) 93

\bibitem{Letessier} J.~Letessier, J.~Rafelski, A.~Tounsi, Phys. Lett. B 328 (1994) 499

\bibitem{BIRO+PESHIER} T.S.~Bir\'o, A.~Peshier, 
	Phys. Lett. B 632 (2006) 247

\bibitem{Broniowski} W.~Broniowski, W.~Florkowski, L.Ya.~Glozman, Phys. Rev. D 70 (2004) 117503

\bibitem{HEAVYION:REVIEW} B.~M\"uller, J.L.~Nagle, 
	Ann. Rev. Nucl. Part. Sci. 56 (2006) 93
	
\bibitem{Muller+Schafer} B.~M\"uller, A.~Sch\"afer, Int. J. Mod. Phys. E 20 (2011) 2235

\bibitem{Muller+Bass} S.A.~Bass, B.~M\"uller, D.K.~Srivastava, Phys. Rev. Lett. 93 (2004) 162301

\bibitem{BMuller} B.~M\"uller, Nucl. Phys. A 750 (2005) 84

\bibitem{Fries+Muller} R.J.~Fries, B.~M\"uller, Eur. Phys. J. C 34 (2004) s279
	
\bibitem{LANDAU:HYDRO} S.Z.~Bilenkij, L.D.~Landau, Nouvo Cim. Suppl. 3 (1956) 15

\bibitem{Khalatnikov} I.M.~Khalatnikov, JETP 27 (1954) 591

\bibitem{HWA} R.C.~Hwa, Phys. Rev. D 10 (1974) 2260 

\bibitem{BJORKEN} J.D.~Bjorken, 
	Phys. Rev. D 27 (1983) 140

\bibitem{OLD:FRANKFURT:Buchwald} 
	G.~Buchwald, G.~Graebner, J.~Theis, J.~Maruhn, and W.~Greiner , Phys. Rev. Lett. 52, (1984) 1594

\bibitem{OLD:FRANKFURT:Rentzsch} 
	T.~Rentzsch, G.~Graebner, J.A.~Maruhn, H.~St\"ocker, W.~Greiner, Z. Phys. C 38 (1988) 237

\bibitem{OLD:FRANKFURT:Waldhauser} 
	B.~Waldhauser, D.H.~Rischke, U.~Katscher, J.A.~Maruhn, H.~St\"ocker, W.~Greiner, Z. Phys. C 54 (1992) 459

\bibitem{OLD:FRANKFURT:Dimitru} 
	A.~Dumitru, J.~Brachmann, E.S.~Fraga, W.~Greiner, A.D.~Jackson, J.T.~Lenaghan, O.~Scavenius, H.~St\"ocker, Heavy Ion Phys. 14 (2001) 121


\bibitem{OLD:FRANKFURT:Gridnev}  
	A.T.~D'yachenko, K.A.~Gridnev and W.~Greiner, J. Phys. G: Nucl. Part. Phys. 40 (2013) 085101

\bibitem{Heinz} 
	P.F.~Kolb, J.~Sollfrank, U.W.~Heinz, 
	Phys. Rev. C 62 (2000) 054909

\bibitem{Heinz1}
	P.F.~Kolb, P.~Huovinen, U.~Heinz, H.~Heiselberg,
	Phys. Lett. B 500 (2001) 232

\bibitem{Heinz2}
	K.S.~Lee, U.~Heinz,
	Z. Phys. C 43 (1989) 425


\bibitem{Romatschke} 
	P.~Romatschke, 
	Class. Quant. Grav. 27 (2010) 025006

\bibitem{Rischke} G.S.~Denicol, T.~Koide, D.H.~Rischke, Phys. Rev. Lett. 105 (2010) 162501

\bibitem{Bouras} 
	I.~Bouras, E.~Moln\'ar, H.~Niemi, Z.~Xu, A.~El, O.~Fochler, C.~Greiner, D.H.~Rischke,
	Phys. Rev. Lett. 103 (2009) 032301

\bibitem{Schenke} 
	C.~Gale, S.~Jeon, B.~Schenke, Int. J. of Mod. Phys. A 28 (2013) 1340011 


\bibitem{Muronga} 
	A.~Muronga, 
	Phys. Rev. C 69 (2004) 034903

\bibitem{Molnar} 
	D.~Moln\'ar, P.~Huovinen, 
	J. Phys. G35 (2008) 104125

\bibitem{CMS}  
	F.~Sikl\'er: {\em private communication}

\bibitem{Muller} 
	T.~Kunihiro, B.~M\"uller, A.~Ohnishi, A.~Sch\"afer, T.T.~Takahashi, A.~Yamamoto, 
	Phys. Rev. D 82 (2010) 114015

\bibitem{Trayanov} 
	B.~M\"uller, A.~Trayanov, 
	Phys. Rev. Lett. 68 (1992) 3387

\bibitem{ChaosBook}  
	T.S.~Bir\'o, S.G.~Matynian, B.~M\"uller: 
	{\em Chaos and Gauge Field Theory}, 
	World Scientific Publishing Co. 1995
	

\bibitem{Strickland} 
	A.~Dumitru, Z.~Nara, B.~Schenke, M.~Strickland, 
	Phys. Rev. C 78 (2008) 024909

\bibitem{Mrowczynski} 
	S.~Mr\'owczy\'nski, 
	Phys. Rev. C 49 (1994) 2191 

\bibitem{UNRUH} 
	W.G.~Unruh, 
	Phys. Rev. D 14 (1976) 870

\bibitem{HAWKING} 
	S.W.~Hawking, 
	Comm. Math. Phys. 43 (1975) 199 

\bibitem{Biro+Schram+Gyulassy} 
	T.S.~Bir\'o, M.~Gyulassy, Z.~Schram,
	Phys. Lett. B 708 (2012) 276

\bibitem{Itzyckson+Zuber} 
	C.~Itzykson, J.B.~Zuber: 
	{\em Quantum field theory}, 
	McGraw-Hill 1980

\bibitem{LandauFiz}  
	L.D.~Landau and E.M.~Lifschitz: 
	{\em Fluid Mechanics}, 
	Pergamon Press 1959

\bibitem{Jackson} 
	D.~Jackson: 
	{\em Classical Electrodynamics}, Wiley, New York, 1975, (Chap 14)

\bibitem{CMSpp} 
	CMS Collaboration, 
	Eur. Phys. J. C 72 (2012) 2164;
	 
	http://ispire.hep.net/record/1123117/hepdata

\bibitem{Jiang} 
	Z.J.~Jiang, Q.G.~Li, H.L.~Zhang,
	Phys. Rev. C 87 (2013) 044902

\end{thebibliography}
\end{document}